# Molecule Induced Strong Exchange Coupling between Ferromagnetic Electrodes of a Magnetic Tunnel Junction

Pawan Tyagi,

Department of Chemical and Materials Engineering, University of Kentucky, Lexington, Kentucky-40506, USA

Current Address: School of Engineering and Applied Science, University of the District of Columbia, Washington DC-20008, USA

**Abstract:** Multilayer edge molecular spintronics device (MEMSD) approach can produce novel logic and memory units for the computers. MEMSD are produced by bridging the molecular channels across the insulator, in the exposed edge region(s) of a magnetic tunnel junction (MTJ). The bridged molecular channels start serving as the dominant exchange coupling medium between the two ferromagnetic electrodes of a MTJ. Present study focus on the effect of molecule enhanced exchange coupling on the magnetic properties of the MTJ. This paper shows that organometallic molecular clusters (OMCs) strongly increased the magnetic coupling between the two ferromagnetic electrodes. SQUID magnetometer showed that OMCs transformed the typical hysteresis magnetization curve of a Co/NiFe/AlOx/NiFe MTJ into linear one. Ferromagnetic resonance studies showed that OMC bridges affected the two fundamental resonance peaks of the Co/NiFe/AlOx/NiFe MTJ. According to magnetic force microscopy, OMCs caused the disappearance of magnetic contrast from the Co/NiFe/AlOx/NiFe tunnel junction area. These three independent and complimentary experiments, suggested the development of extremely strong interlayer exchange coupling. This work delineated a practical route to control the exchange coupling between ferromagnetic electrodes. Ability to tailor magnetic coupling can lead to the development of molecule based quantum computation device architecture.

**Introduction:** Molecular spintronics devices (MSDs) are promising to produce new computer logic and advanced spintronics devices. According to theoretical and experimental studies, MSDs are shown to possess several new remarkable attributes which are not observed with conventional magneto resistance devices. MSDs' development will critically depend on the ability to distinguish the role of molecules in bringing out new device attributes. Conventionally, spintronics devices are investigated by performing highly correlated transport and magnetic studies. Magnetic characterizations have played crucial role in the advancement of magnetic tunnel junction (MTJ) based spin valves [1-3]. Similar magnetic characterizations will be even more crucial for the development of MSDs. Magnetic characterizations are required to gauge the effect of molecule(s) on the magnetic coupling between two ferromagnetic (FM) electrodes [4]. It is due to the fact that in a MSD molecules are not merely the spin transport channels; molecules are capable of strengthening the inter-electrode exchange coupling [5]. For the study of molecules' role in a MSD, a test platform that can be studied before and after attaching the molecular channels between two electrodes will be highly desirable [6-8]. Tunnel junction based multilayer edge molecular spintronics devices (MEMSD) are the promising candidate for the development of novel MSDs. MEMSD enables the insightful magnetic characterizations to study the effect of molecular channels in producing new device attributes [8]. MEMSDs are produced by bridging the molecular channels across the insulator of a MTJ, in the exposed edge region (Fig. 1). As a major advantage, a MTJ can be easily characterized before transforming it into a MEMSD.

Effectiveness of molecule as an exchange coupler can be better understood by the evaluation of analogous entities like impurities and nanoclusters. Several research groups showed that MTJs experienced significant increase in the exchange coupling due to impurities [9] and nanoclusters [10]. Utilization of molecules can serve the role of nanoclusters and impurities to tailor the inter-electrode exchange coupling and yield mass producible devices. Molecules are much more reproducible than the impurities and nanoclusters. Precise control over molecule's size, spin [11] and spatial location can lead to an extremely strong inter-electrode exchange coupling and unprecedented advancement in the field of MSDs [6, 12-16]. If these coupling become strong beyond a limit [17] then FM electrodes may undergo dramatic changes [18] and can lead to anomalous phenomenon [5]. Magnetic characterizations are essential to investigate the effect of molecule's induced exchange coupling on the MSDs properties. This paper investigated the effect of molecules on the magnetic properties of a MTJ based MEMSD (Fig. 1a-b). In our studies organometallic molecular clusters (OMCs) [8, 19] were utilized to tailor the exchange coupling between the FM electrodes (Fig.1c). Super conducting quantum interference device (SQUID) magnetometer, ferromagnetic resonance (FMR), and magnetic force microscopy (MFM) were used for the magnetic characterization. Here we present magnetic studies which showed a strong effect of molecules on the exchange coupling between two FM electrodes of the magnetic tunnel junction. The magnetic studies are utilized in explaining the dramatic current suppression observed with the MEMSD systems.

**Experimental details:** MEMSDs for the magnetic characterizations were designed so that to: (a) extraneous magnetic electrodes, away from the molecular junction, do not mask the OMCs effect (b) a chip studied by the magnetic characterizations produced unequivocal response before and after the treatment with OMCs. To achieve these two objectives MEMSDs were produced in the form of cylindrical dots, with a typical MTJ configuration and exposed side edges (Fig.1a-b). Cylindrical dot geometry with a 3-5 $\mu$m diameter allowed OMCs (Fig. 1c) to bridge across the insulator of a MTJ along its perimeter. A typical sample used for the magnetic characterizations contained 7000-21000 MTJs. The inclusion of a large number of MTJ cylindrical dots ensured the reasonably strong magnetic signal in a variety of magnetic characterizations. Fabrication of the MEMSD dots required the creation of cylindrical cavities in a photoresist layer by photolithography. Photoresist layer was produced by spin coating Shipley 1813 photoresist at 3000 rpm speed on a thermally oxidized silicon chip. After spin-coating, photoresist was backed at 90-100 ºC for 1-2 minutes. Hardened photoresist was exposed to UV light for ~30 seconds with Karl Suss Mask Aligner. Subsequently, photoresist was developed in Shipley MF 150 developer for ~40-60 seconds. After this step, a typical chip contained a large number of cavities for the simultaneous deposition of cylindrical MTJs. To fabricate MTJ cylinders, sequentially tantalum (Ta), cobalt (Co), 80% nickel-20% iron alloy (NiFe), alumina (AlOx), and NiFe films were sputter deposited in the photoresist cavity molds (Fig. 1d). Liftoff step resulted in an array of MTJ dots with ~4 $\mu$m inter-dot separation and Ta[5nm]/Co[5nm]/NiFe[5nm]/AlOx[2nm]/NiFe[12 nm] configuration. To measure the magnetic response before OMCs attachment MTJs were subjected to numerous magnetic studies.

In order to transform MTJ into MEMSD, OMCs were electrochemically bridged across the insulator of a tunnel junction (Fig. 1b). For molecule attachment, MTJs were submerged in the 2 mM OMC solution in dichloromethane. Then two electrical probes were dipped in the drop of OMC solution, and ±100 mV alternating voltage with a time interval of 0.02 seconds was applied. This step presumably de-protected the thiol functional groups present at the end of four alkane tethers (L) of an OMC (Fig. 1c) [20]. Those OMCs whose two alkane tethers bridged across the AlOx insulator of a MTJ (Fig.1b), become the part of molecular device and yielded a MEMSD cylinder. After the molecule attachment step MEMSDs were sequentially rinsed in di choloro methane, iso-propyl alcohol, and deionized water. Finally, samples were again subjected to magnetic characterizations to measure the effect of OMCs on MTJs' basic magnetic properties.

Magnetic properties of a MEMSD were studied using three measurement techniques: (a) magnetization study, (b) ferromagnetic resonance (FMR), and (c) magnetic force microscopy (MFM). The magnetization studies were performed using the Quantum Design MPMS XL 7T SQUID magnetometer. MEMSD samples, mounted on a glass rod with the help of epoxy, were introduced into the helium gas flushed superconducting coil of SQUID magnetometer. Samples were carefully centered in magnetometer sample cavity to get the optimum signal intensity. The temperature during magnetization study was kept at 150 K. Very low temperatures in 4 K range produced detrimental thermal stresses which damaged the tunnel barrier. At higher temperature close to room temperature, typical magnetization data was highly noisy.

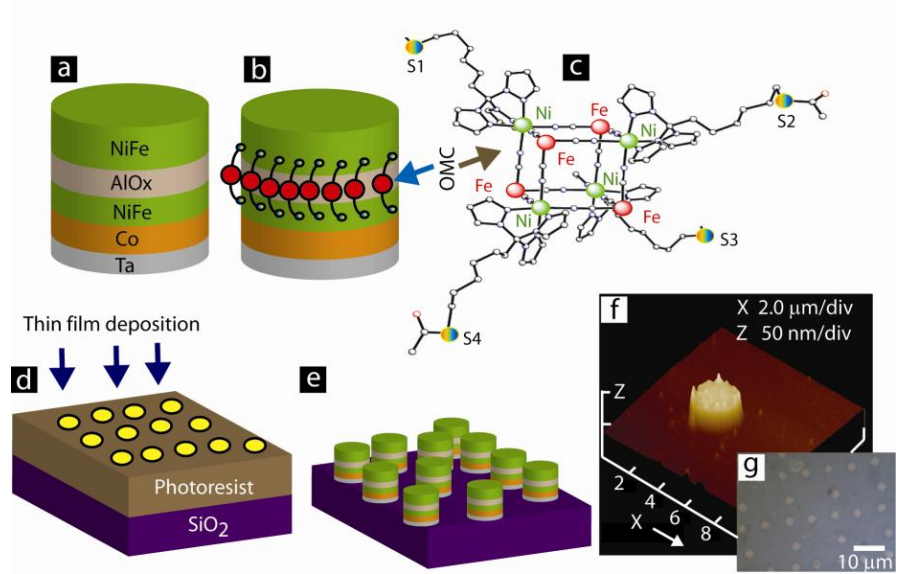

Fig. 1: MEMSD cylinders for magnetic characterizations: Schematic of MTJ (Ta/Co/NiFe/AlOx/NiFe) (a) before and (b) after bridging of OMC channels across AlOx. (c) Chemical structure of an OMC unit, sulfur (S) helped covalent bonding of OMCs to the NiFe electrodes. Fabrication of MEMSD cylinders required (d) the creation of several thousand cavities in the photoresist layer on insulating substrate, and (e) the sequential sputter deposition of all the layers through the cavity. The lift-off step produced several thousand MTJs. OMC attachments convert each MTJ into (b) MEMSD. (f) 3D AFM image of a MEMSD. (g) Optical micrograph of a section of the chip containing ~7000 MEMSDs.

For FMR studies Brucker EMX EPR spectrometer equipped with Brucker Mirowave Bridge ER 041MR and Brucker Power Supply ER 081(90/30) was utilized. For all the experiments ~9.7 GHz microwave frequency and room temperature were maintained. A magnetic field was applied in the sample plane to study the uniform modes of thin films and multilayers. Application of the magnetic field perpendicular to the sample plane did not produce promising distinct FMR signals, for the meaningful study of OMCs effect on MTJ. Before every FMR study, empty cavity's spectra was checked for the background signal at 5 fold higher gain than that used for the typical MTJ and MEMSD samples.

MFM studies were performed using Digital instrument's multimode AFM. Highly sensitive MFM tips (supplied by Nanosensors.Inc) with the following specifications were utilized: Type PPP-MFMR, tip side Co coated, force constant in 0.5-9.5 N/m range, resonance frequency 45-115 kHz. During MFM scans, the gap between the probe's tip and substrate was 10-150 nm. For the high resolution and minimally interrupted MFM studies a 100 nm tip sample gap was frequently utilized.

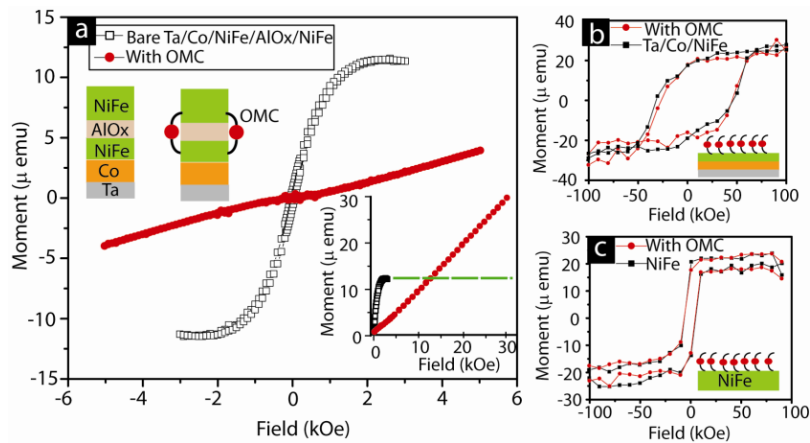

Fig. 2: Magnetization study of MEMSD: (a) magnetization curve of (a) Ta/Co/NiFe/AlOx/NiFe MTJ before and after OMCs attachment. Inset show OMCs induced magnetization curve did not saturate up to 30 kOe (~3 T). OMCs did not affect the magnetization of individual (b) bottom (Ta/Co/NiFe) and (c) NiFe top electrode.

**Results and discussion:** Initially, magnetization studies were performed to elucidate the effect of OMCs on the MTJ and overall MEMSD's attributes. During the magnetization studies magnetic field was applied in the sample plane. The first sample, which was subjected to magnetization study, was also studied by FMR and MFM along with other MEMSD samples; hence, for easy identification we named it as MEMSD-A. The MTJ (Ta (5 nm)/Co (5 nm)/NiFe (5 nm)/AlOx (2 nm)/NiFe (12 nm)) used for MEMSD-A exhibited a saturation magnetization value of ~1.0 × $10^{-4}$ emu at ~2 kOe. For the comparison of MTJ's saturation field with that of individual electrode's saturation field top and bottom electrodes were also subjected to the magnetization study. The saturation magnetization for the unpatterned top [NiFe (12 nm)] and bottom [Ta(5nm)/Co(5 nm)/NiFe (5 nm)] electrodes were ~10 Oe and ~60 Oe, respectively (Fig 2 b-c). A higher coercive field for MTJ dots, as compared to the unpatterned individual electrodes is supposedly due to the shape anisotropy of a MEMSD cylinder and inter-dot dipolar interactions [21].

Table 1: Parameters for the calculation of OMCs induced antiferromagnetic coupling strength.

| Parameters | Values |
|---|---|
| **Magnetization of NiFe film(Ma)** | $1.2 \times 10^5$ A/m |
| **Thickness of soft film (t)** | $1.0 \times 10^{-8}$ m |
| **Applied saturation field ($H_{ext}$)** | 3T |
| **MEMSD dot diameter** | $3.0 \times 10^{-6}$ m |
| **Area** | $7.1 \times 10^{-12}$ m$^2$ |
| **Volume/MEMSD cylinder** | $2.0 \times 10^{-19}$ m$^3$ |
| **No. of molecules** | $1.0 \times 10^4$ |
| **Antiferromagnetic coupling energy** | $1.5 \times 10^{-6}$ erg |
| **Energy per unit junction area** | 21.87 erg/cm$^2$ |
| **Energy per molecule (in erg)** | $2.2 \times 10^{-3}$ erg/(cm$^2$ –molecule) |

The OMCs radically affected the magnetization of Ta/Co/NiFe/AlOx/NiFe MTJ. OMCs bridging reduced the MTJ's magnetic moment by ~90% and produced a linear magnetization curve (Fig. 2a). In essence, OMCs transformed MTJ hysteresis curve into a linear magnetization curve. A linear magnetization response is the characteristics of paramagnetic materials or two antiferromagnetically coupled FM electrodes [22, 23]. The MEMSD configuration, comprising a MTJ with two FM electrodes and OMCs, support the hypothesis that OMC induced linear magnetization is due to the development of strong antiferromagnetic coupling (AFC) between two FM electrodes [22, 24]. A system comprising two FM electrodes showed a linear magnetization curve when AFC strengths (*J*) was more than the product of the anisotropy constant (*K*) and FM electrode thickness (*t*) [22], i.e., *J* > *Kt*. A linear magnetization curve can be used for the calculation of AFC strength. In a related study, a MTJ with iron/silicon/iron configuration produced the linear magnetization curve for the limited range of magnetic field [24], after certain magnetic field the linear magnetization saturated. Magnitude of the saturation field for an AFC induced linear magnetization curve, is believed to overcome the inter-electrode exchange strength between the two FM electrodes. Based on the saturation magnetic field exchange coupling strength can be expressed by the following expression:

$$J = V.M_a.H_{ext} \tag{1}$$

Here, V= volume of FM electrodes, Ma= saturation magnetic field, and $H_{ext}$ is the magnetic field at which magnetization saturate. In order to utilize this expression to calculate the strength of OMC induced inter-electrode AFC we need to know $H_{ext}$, external magnetic field at which magnetization saturated. However, magnetization of MEMSD-A did not saturate up to 3 T external magnetic field (inset of Fig. 2a); due to the lack of resources we were unable to raise $H_{ext}$>3 T. Assuming that in our case H is at least 3 T, we calculated the magnitude of J. Details of calculation J are tabulated in Table 1.

*It is noteworthy that the magnitude of the estimated OMC induced J is consistent with the magnitude of extrapolated J from the analogous systems [21, 22].* A system analogous to MEMSD and with ferromagnet/nonmagnet/ferromagnet system with 1.54 erg/cm$^2$ AFC strength, showed the saturation of linear magnetization curve at ~0.5 T [22]. We did not observe saturation of OMC induced magnetization loop up to 3 T, and hence we assumed that 3T to be at least the lower bound of saturation field for our MEMSD system. Considering that the MEMSD's linear magnetization loop saturated at 3 T, the linear extrapolation of literature data [22] yielded corresponding OMCs induced AFC strength to be 9 erg/cm$^2$. This magnitude of extrapolated J (9 erg/cm$^2$) is of the same order of the J directly estimated by using MEMSD specific parameters (21.87 erg/cm$^2$). In summary, ~10,000 OMC enhanced the AFC between two FM electrodes of the Co/NiFe/AlOx/NiFe MTJ to 21.87 erg/cm$^2$. Enhancement of inter-FM electrode coupling per OMC unit is ~2.2 x 10$^{-3}$ ergs/cm$^2$, which is close to the inter-electrode coupling strength of a typical bare MTJ [9].

Molecules [25], like quantum dots [26] and atomic defects [9], enable a stronger exchange coupling between the two FM electrodes. Strength of molecular coupling increases with their population. In the present case, cumulative effect of ~10,000 OMCs becomes strong to the extent that magnetization properties of the Co/NiFe/AlOx/NiFe MTJ changed dramatically. Analogous to MEMSD, Ni breakjunction with C$_{60}$ magnetic molecule(s) [5], also produced unprecedented exchange coupling between two Ni FM electrodes. In both the cases magnetic molecules showed dramatic effect when simultaneously coupled to the two FM electrodes. *Can*

*OMCs interaction with the single FM electrode produce the dramatic change in its magnetic properties?* To investigate the answer to this question we studied unpatterned FM electrodes before and after treating them with OMCs (Fig. 2 b-c). Covalent bonding of OMCs to unpatterned Ta/Co/NiFe (bottom electrode material) and NiFe (top electrode material) did not produce any noticeable change in magnetic moment (Fig. 2 b-c). This study established that the interaction between OMCs and single FM electrode is unable to produce the same dramatic effect, which it produced when simultaneously interacting with the two FM electrodes. In order to ascertain that observed linear magnetization curve on MEMSD-A was due to development of strong AFC, we performed magnetic moment versus temperature study. If MEMSD's FM electrodes are antiferromagnetically coupled then a change in cumulative magnetic moment must occur when the supplied thermal energy overcome the OMCs induced inter-electrode coupling energy [27].

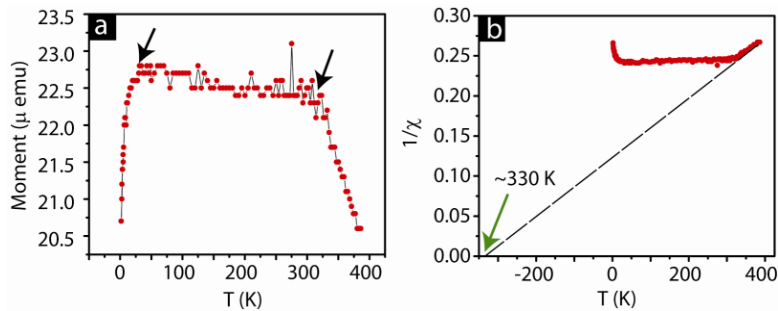

Fig. 3: (a) Temperature versus magnetic moment, and (b) inverse of magnetic susceptibility versus temperature for the MEMSD-A.

In our experiment, magnetic moment first increased up to ~30 K; presumably, increasing temperature enabled the rotation of rather stiff and randomly oriented magnetic domains to cause increase in magnetic moment (Fig. 3a). The second transition around 330 K caused sharp reduction in the magnetic moment. We argue that heating the sample to 330 K disrupted the OMCs induced AFC between the two FM electrodes of the Ta/CoNiFe/AlOx/NiFe MTJ. To determine the nature and magnitude of OMCs induced coupling on MEMSD-A, we plotted the inverse of magnetic susceptibility as a function of temperature (Fig. 3b). The intersection point of a tangent on the part after the second transition produced the useful information (Fig. 3b). This tangent intersected temperature axis at -330 K [27]; according to Curie-Weiss law this intersection on negative temperature axis indicates the existence of AFC [27], and represented the Neel temperature. However, we are still unsure if the magnitude of the OMC induced AFC is ~330 K. Conventional theories are unexpected to reveal the complete understanding of OMCs effect. OMCs can also produce a second order biquadratic coupling [28] and spin fluctuation induced couplings [5]. These effects are not encompassed by the conventional method of calculating AFC strength using moment versus temperature study [27]. To further understand the OMCs induced strong coupling same MEMSD-A sample was also studied by FMR and MFM; results of these studies are discussed elsewhere in this paper.

We found that OMCs effect was prominent when a test bed MTJ had two FM electrodes. *Can OMCs also affect the magnetic moment of a tunnel junction with one FM electrode?* To investigate the OMCs effect two MTJ configurations with Ta/Co/NiFe/AlOx/Pd(palladium), and Pd/AlOx/NiFe, were prepared. These configurations involved top or bottom electrodes of the Ta/Co/NiFe/AlOx/NiFe MTJ configuration, on which OMCs showed the dramatic effect. OMCs produced highly intriguing effect on MTJ with one FM electrode. OMC increased the magnetic moment of MTJ with Ta/Co/NiFe bottom FM electrode (Fig. 4a). On the other hand, OMCs decreased the magnetic moment of MTJ with NiFe FM electrode (Fig. 4b). These magnetization studies (Fig. 4) suggested that OMCs interaction is specific to the FM electrode. Present study emphasized that a vast variety of MEMSDs can be produced by having a large combination of magnetic molecules and FM electrodes. By now we have shown three independent samples which showed OMC induced changes in magnetic moment. Change in magnetic moment can impart dramatic changes in the transport characteristics of the MEMSD [1].

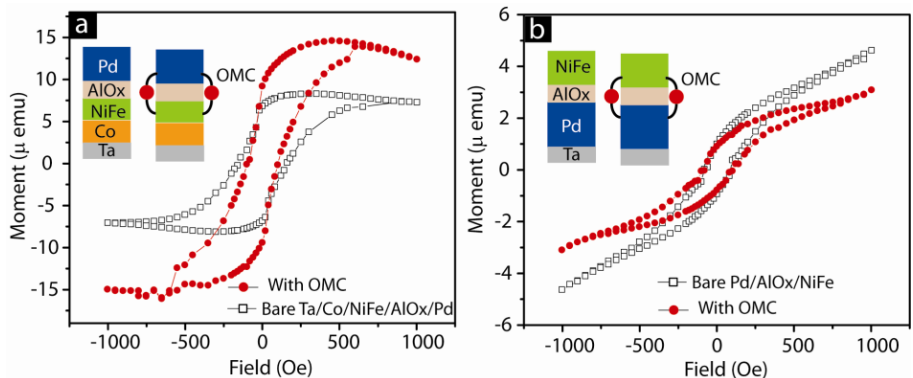

Fig.4: Effect of OMCs on the magnetization of MTJ with single (a) Ta/Co/NiFe and (b) NiFe FM electrode.

Does OMCs effect depend on the nature of preexisting coupling between the two FM electrodes of a MTJ? This question arise from the fact that development of OMCs induced exchange coupling was observed on the MTJ with preexisting weak AFC (Fig.1) To further probe the nature of OMC induced coupling we also studied MTJs with the ferromagnetically coupled FM electrodes. As discussed elsewhere in this paper FMR studies were performed to determine the nature of coupling prior to OMC attachment. The Ta/Co/NiFe/AlOx/NiFe configuration possessed weak AFC. Bridging of OMCs dramatically enhanced the AFC strength (Fig. 3). Interestingly, just small modification in Ta/Co/NiFe/AlOx/NiFe produced MTJ with ferromagnetic inter-electrode coupling. The addition of Ta (5 nm) on the top of Ta/Co/NiFe/AlOx/NiFe MTJ yielded Ta/Co/NiFe/AlOx/NiFe/Ta MTJ configuration with inter-electrode ferromagnetic coupling. Interestingly, OMCs dramatically enhanced the magnetic moment of the Ta/Co/NiFe/AlOx/NiFe/Ta MTJ (Fig. 5). *OMCs induced magnetic moment increase was equivalent to the deposition of additional magnetic material*. Our magnetization studies (Fig. 2 and 5) exhibited that OMCs enhanced the strength of preexisting inter-electrode coupling. Another utility of the magnetization study reported in Fig. 5 is to serve as a complimentary experiment to support the magnetization study on MEMSD-A (Fig. 2a). Same OMCs which dramatically increased the magnetic moment of MTJ with inter-electrode ferromagnetic coupling (Fig.5), also dramatically reduced the magnetic moment of the MTJ with interelectrode antiferromagnetic coupling (Fig. 2a). It is apparent that, OMCs are only responsible for the dramatic response from MTJs. It is also noteworthy that four MEMSDs utilized in magnetization studies (Fig. 2-5) were produced using identical experimental conditions. Additionally, the OMC batch [19] utilized in magnetization studies also produced dramatic current suppression [4] and successful multilayer edge molecular electronics devices [8].

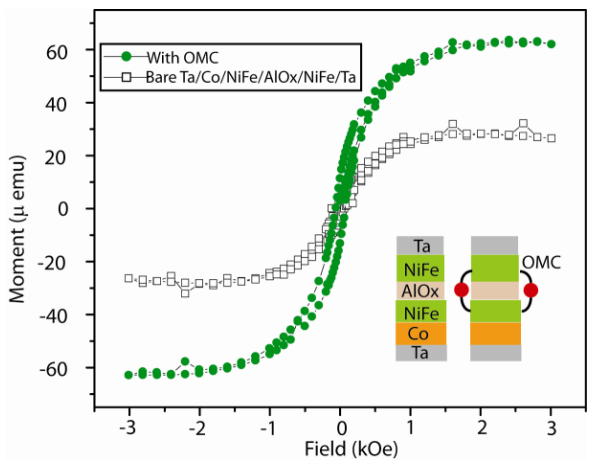

Fig. 5: OMCs increased the magnetic moment of MTJ (Ta/Co/NiFe/AlOx/NiFe/Ta), showing weak ferromagnetic coupling between the two FM electrodes in bare state.

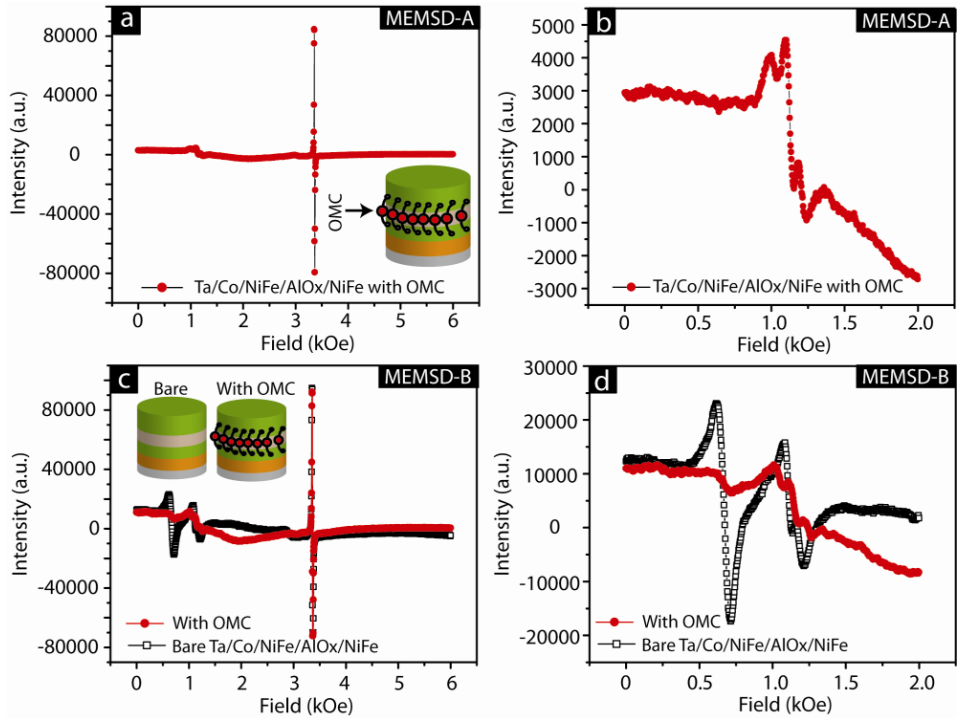

Fig. 6: FMR study of Ta/Co/NiFe/AlOx/NiFe MTJ based MEMSD: FMR spectra of MEMSD-A in (a) 0-6 kOe and (b) 0-2 kOe range. (c) FMR spectra of a MTJ before and after transforming it into MEMSD (named as MEMSD-B). (d) FMR spectra zoomed in 0-2 kOe range.

FMR characterization was utilized to investigate the validity of magnetization studies performed on MEMSDs. First FMR study was carried out on MEMSD-A, the sample which showed dramatic OMCs effect during magnetization study (Fig. 2 and 3). FMR response from MEMSD-A showed a major peak at ~3.4 kOe and a very small hump around ~1.2 kOe (Fig. 6a). The sharp peak at ~ 3.4 kOe was due to the carbon tape, used for sticking the sample on to the carrier rod. FMR peak for the carbon tape was used for monitoring the variation in experimental conditions. The small hump around ~1.2 kOe is due to the MEMSD-A. A typical MTJ showed two FMR peaks. However, the observation of only one peak with MEMSD-A is in agreement with the FMR model developed for analogous system of strongly coupled FM electrodes [17, 29]. FMR model exhibited that when inter-electrode AFC strength exceed a critical value then only single FMR mode will appear [17]. Moreover, the intensity of MEMSD-A's single FMR mode was much weaker than the intensity of typical FMR modes of the Co/NiFe/AlOx/NiFe MTJ (Fig. 6b). The reduction in intensity can be explained in terms of the following FMR intensity (I) expression [29]:

$$I = \frac{2MVF(H - H_{Keff})}{2H - H_{Keff}} \qquad (2)$$

Where, M is magnetization, V is sample volume, and H is external field. F is the proportionality factor, which included strength of FMR experimental factors like RF magnetic field, microwave frequency; lock in amplifier operational conditions etc. $H_{Keff}$ is effective anisotropy field and given by $H_{Keff}$= (2K/M)-4πM, where K is the magnetocrystalline anisotropy [29].

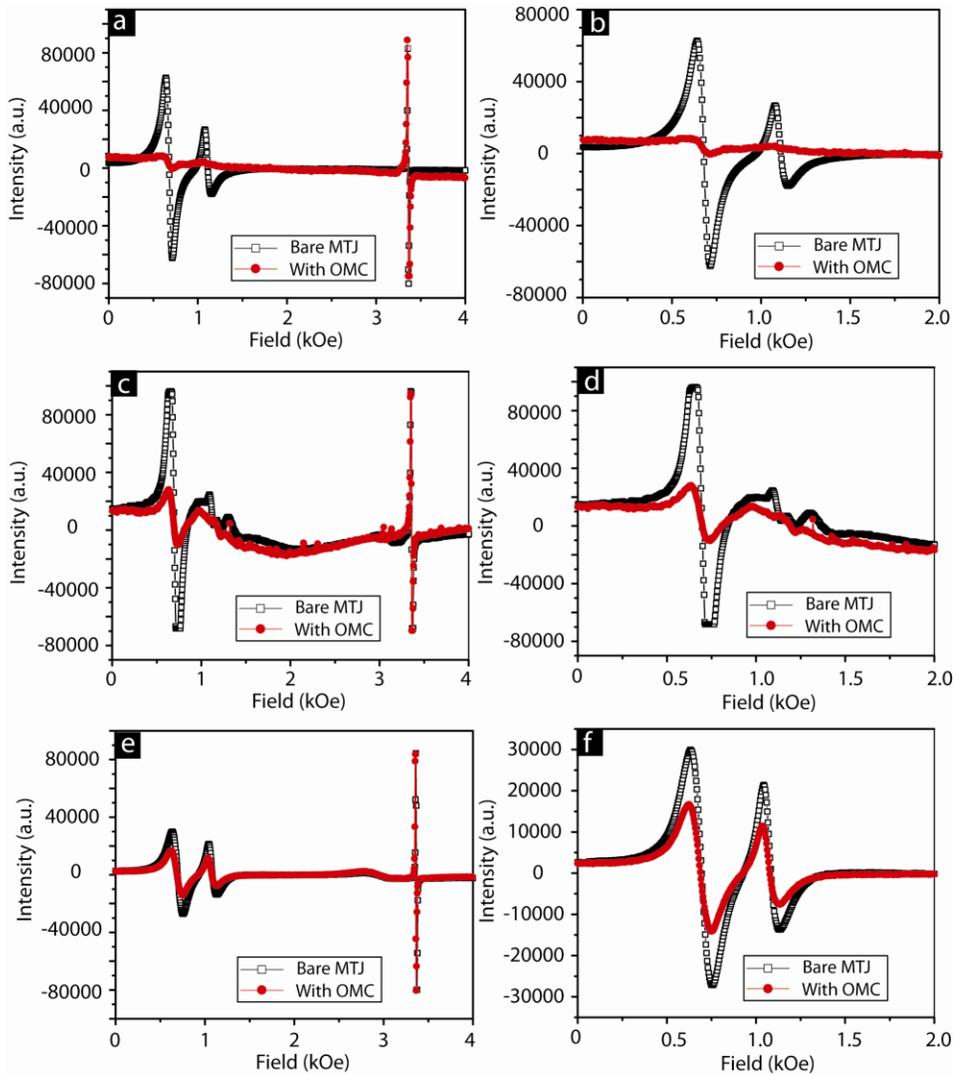

Fig. 7: OMCs induced FMR mode suppression on (a) MTJ with ~2 nm AlOx and produced in optimized experimental conditions, (b) MTJ with ~1 nm thick AlOx, and (c) MTJ treated with a new batch of molecules and deviated experimental factors.

According to the equation (2), the intensity of FMR mode depended on the sample magnetization M. Our magnetization studies on MEMSD-A (Fig. 2a) showed that OMCs reduced the magnetization of Ta/Co/NiFe/AlOx/NiFe MTJ by ~ 90% in 0.5-2 kOe range. Hence, the observation of weak intensity of single FMR mode (Fig. 3a-b) is in agreement with the magnetization study on the same MEMSD-A sample (Fig. 2a). To further strengthen the FMR study on MEMSD-A, we attempted to compare the FMR response from the same MTJ before and after transforming it into a MEMSD; the MTJ for MEMSD-A could not be studied prior to OMCs treatment because FMR studies were planned several months after the magnetization studies (Fig. 2a). The MTJ utilized in the second FMR study was transformed into a MEMSD and named it as MEMSD-B.

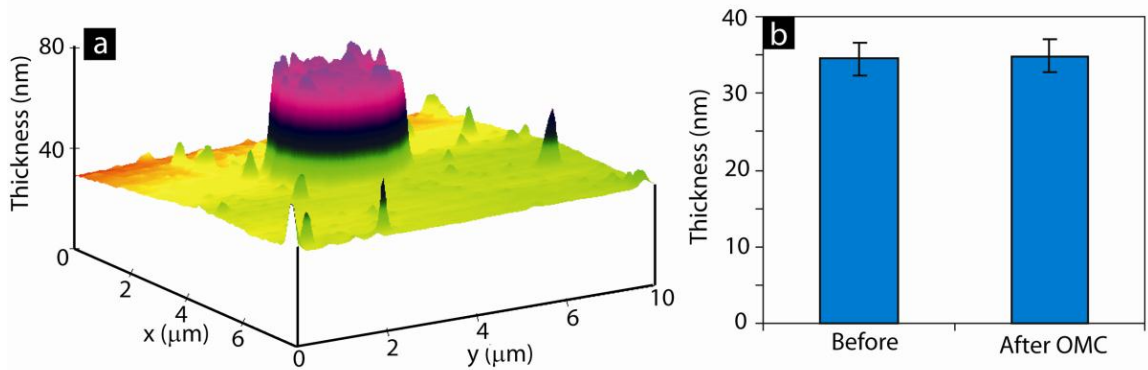

Fig. 8: AFM imaging of MEMSD showing FMR modes suppression: (a) 3D image of a MEMSD and (b) statistical comparison of MTJ cylinder height before and after transforming it into MEMSD.

FMR spectra of the bare Ta/Co/NiFe/AlOx/NiFe MTJ, utilized for MEMSD-B, exhibited acoustic mode (peak with higher intensity) at ~670 Oe and optical mode (peak with smaller intensity) at 1120 Oe. The position of optical mode after the acoustic mode imply an antiferromagnetic coupling [30] between the two FM electrodes via ~2 nm AlOx insulator. We also compared the two modes of bare MTJ with the FMR modes recorded on the NiFe and TaCoNiFe electrodes deposited in isolation. MTJ's acoustic mode (optical mode) was close to the mode position of isolated and separately grown NiFe (Ta/Co/NiFe) layer; the difference between acoustic mode (optical mode) position and isolated NiFe (TaCoNiFe) mode position was 40-70 (30-50) Oe. After the bridging of OMCs, MTJ transformed into MEMSD-B and experienced dramatic change in the intensity of its two FMR peaks (Fig. 6c-d). Bridging of OMCs across AlOx insulator (inset of Fig. 6c) diminished the acoustic mode at 0.67 kOe (Fig.

6d). Only a hump around the optical mode position remained (Fig. 6d), which is consistent with the FMR spectra of MEMSD-A (Fig. 6b). The appearance of only a weak FMR peak on MEMSD-B represented the development of OMCs induced strong coupling between two FM electrodes [17]. The FMR result on MEMSD-B agrees with that of MEMSD-A. It is noteworthy that FMR peak intensity for the carbon tape remained unchanged during the two experiments, signifying that the parameters used in FMR studies before and after OMC treatment of the MTJ were the same (Fig. 6d).

FMR studies on several MEMSDs supported the OMCs induced dramatic change in magnetic coupling. In a control study a Ta/Co/NiFe/AlOx/NiFe MTJ experienced the OMC induced FMR mode suppression (Fig. 7a-b). However, OMCs effect varied among control MEMSD samples (Fig. 7). Variation in the experimental conditions, molecule quality, photolithography conditions produced partial OMCs effect in several cases (Fig. 7c-f). The AlOx sturdiness was the key parameter for invoking full potential of OMC bridges. For instance, the MTJ sample where AlOx thickness was reduced to ~1.0 nm OMCs yielded partial effect (Fig. 7c-d). Presumably, attempt to produce thinner AlOx barrier lead to the production of defective or leaky MTJs, which failed to show complete FMR mode suppression. Two FMR modes of the bare MTJ with ~1nm AlOx (Fig. 7b) were not as clear as observed with thicker AlOx (2 nm) tunnel barrier (Fig. 7a). Another control MTJ showed partial OMCs induced mode suppression (Fig. 7e-f) with the OMCs batch. This partial mode suppression commenced with unsound OMCs. The same batch of OMCs was also ineffective in yielding conclusive multilayer edge molecular electronics devices [8].

To ensure that severe damage, like complete etching of MTJs cylindrical dots, is not the reason behind FMR mode disappearance, several physical characterizations were performed. For instance, the control MTJ sample which experienced strong OMCs' effect (Fig. 7a) was subjected to AFM thickness measurement before and after OMC treatment. More than 140 MTJ cylindrical dots were studied from the various parts of the control sample before and after OMCs attachment. During this study, topography and 3D profile of individual MTJ cylinders was recorded (Fig. 8a). Thickness measurement by the AFM showed unnoticeable statistical difference in the height of cylindrical dots (Fig. 8b) proving that etching type physical damage did not occur. Similar physical thickness before and after OMC treatment also evidenced that oxidation of FM electrodes, which could have caused volume expansion, did not take place. Additionally, high magnification optical microscopy with polarized light was extensively employed to monitor the sample integrity. In some cases, a small population of dots (5-10%) underwent partial to complete damage during the electrochemical OMC attachment step. FMR study of such samples did not exhibit more than 10% change in signal intensity after OMC treatment. A control sample, which was unknowingly processed with deviated photolithography parameters, underwent accidental wash off of > 20% MTJ cylindrical dots during OMC treatment. This sample represented the case of highly damaged MEMSDs, and hence this sample was utilized to study the FMR response in the worst-case scenario. This sample produced strong FMR modes (Fig. 9a), which were quite different from the stable sample (Fig. 6). The optical micrographs collected from the different region of this damaged sample showed that a number of MTJs remained in the sound state (Fig. 9b-c).

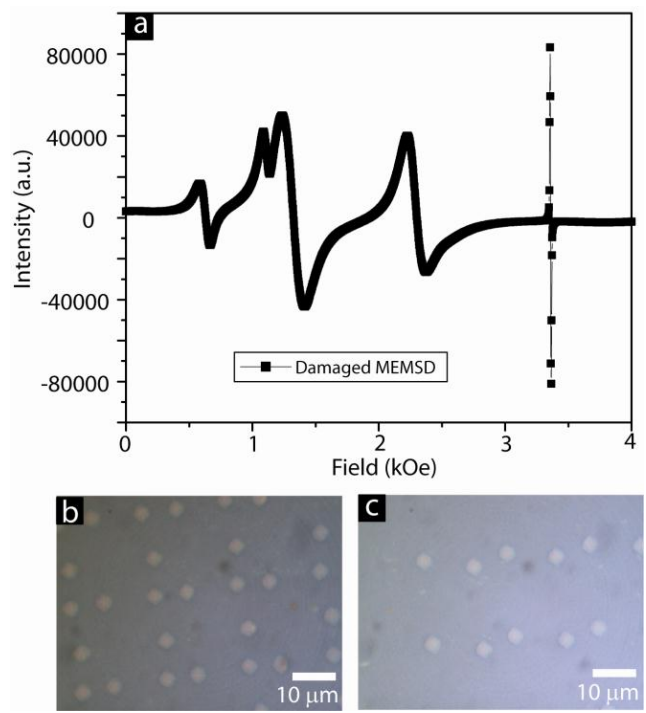

Fig. 9: FMR study of the damaged MEMSDs showing partial or complete damage to ~30% MTJ cylinders during electrochemistry for OMCs attachment. (b) and (c) are the optical micrographs from two different regions of the damaged MEMSD chip.

To investigate the potential damage from the electrochemical OMC attachment step two additional control experiments were performed. In the first control experiment MTJ cylindrical dots were prepared with ~ 4 nm AlOx. Use of ~4 nm thick AlOx disabled the bridging of ~ 3 nm OMCs across AlOx insulator. In this case OMCs were not expected to enhance the exchange coupling between the FM electrodes of MTJ. FMR studies before and after OMC treatment showed that OMCs did not produce any significant change in the intensity of FMR modes (Fig.

10a). In the second control experiment, cylindrical dots were made up of only Ta/Co/NiFe bottom electrode materials. The significance of this control experiment is that OMC solution can access the edges of Ta/Co/NiFe cylindrical dots and to etch-sensitive Co therein. FMR studies before and after OMC treatment showed no change in FMR modes' intensity (Fig. 10b). This observation is consistent with our control experiments in which Co etching rate was <10 nm/OMC treatment. A 3 µm diameter dot is unexpected to loose significant Co to show measureable change in FMR mode intensity.

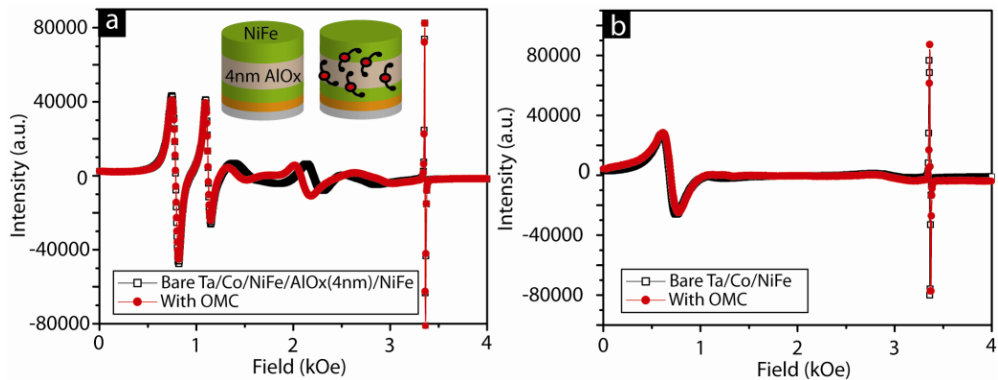

Fig. 10: Monitoring chemical damage due to electrochemical OMC treatment step: (a) FMR response of MTJs with ~ 4 nm AlOx and (b) FMR study of only bottom FM electrode that contained etch sensitive Co.

Can unnoticed oxidation of NiFe produce antiferromagnetic oxides like NiO and FeO, thus leading to the disappearance of FMR modes? To check this possibility a NiFe electrode was purposefully plasma oxidized to reveal the nature of changes in the FMR mode. Plasma oxidation of NiFe did not reduce the intensity of FMR mode recorded before oxidation. Interestingly, additional resonance peak emerged at 800 Oe [4]. We attributed this additional peak to the NiFe surface-oxides.

We also investigated the effect of sample misalignment during FMR study on the mode intensity. To check this possibility a chip with Ta/Co/NiFe/AlOx/NiFe MTJ cylindrical dots was studied. This sample was intentionally misaligned by ~ 10⁰ with respect to the direction of external magnetic field. We were able to precisely place the sample in the plane of applied magnetic field with an accuracy of ±10⁰. Two major FMR modes remained unchanged with small variation (Fig. 11). We also studied the effect of reinserting the sample in FMR measurement cavity. Reinsertion of sample in FMR cavity was essential to do the comparative study before and after OMC treatment. No measurable change in FMR mode intensity was observed when sample was reinserted (Fig.11). Additionally, FMR mode for the carbon tape at ~ 3.3 kOe remained unchanged with this misalignment and reinserting of the sample. In addition to major peaks FMR spectra also showed the presence of labile walker modes between ~1.3 to 3.3 kOe. However, these secondary effects were not utilized to gauge the OMCs effect, and their study is beyond the scope of present work.

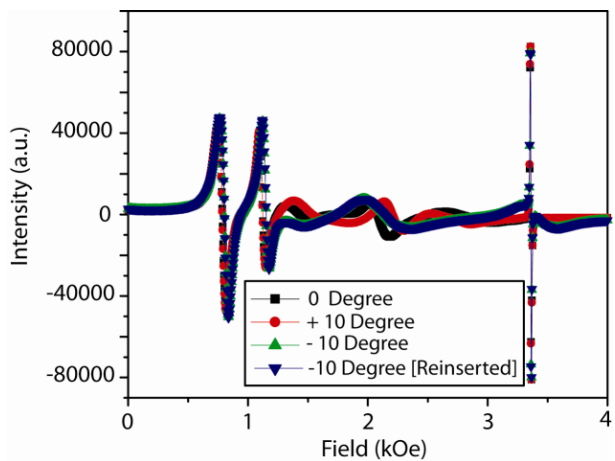

Fig. 11: Effect of variation in angle of sample orientation with respect to direction of the applied magnetic field.

We observed that OMCs effect was mainly observed on the Ta/CoNiFe/AlOx/NiFe MTJs with inherent weak AFC. Does OMCs effect depend on the initial coupling between FM electrodes of a bare MTJ, i.e. before OMC treatment step? To answer this question we utilized a MTJ with ferromagnetic inter-electrode coupling. Fortunately, a MTJ with ferromagnetic inter-electrode coupling was realized with the minor modification in the thin film configuration showing AFC in the bare state. In a serendipitously discovery we found that just the addition of Ta layer above the top NiFe electrode produced Ta/CoNiFe/AlOx/NiFe/Ta MTJ, which possessed the ferromagnetic coupling between the two FM electrodes. For Ta/CoNiFe/AlOx/NiFe/Ta MTJ the low intensity FMR mode (optical) occurred at lower magnetic field while high intensity FMR mode (acoustic) occurred at higher magnetic field. This particular characteristic of FMR spectra represents the appearance of ferromagnetic coupling between the two FM electrodes [21]. For the present FMR study we used MEMSD sample which also showed unambiguous OMCs effect in the magnetization study (Fig. 5). FMR spectra of OMCs treated Ta/CoNiFe/AlOx/NiFe/Ta MTJ produced only one prominent mode (Fig. 12 a). The reasonably high FMR mode intensity of this single mode is consistent with the enhanced magnetization on the same MEMSD (Fig. 5). In the same manner, very low intensity of single FMR mode on MEMSD-A, utilizing MTJ with antiferromagnetically coupled FM electrodes, was consistent with the reduced magnetization on this sample (Fig. 2a). According to theoretical modeling the development of single FMR mode is indicative of strong ferromagnetic coupling between the two FM electrodes [17]. Two FM electrodes coupled by the strong ferromagnetic coupling behaved like a single ferromagnetic electrode and produced a single mode [17]. We were unable to compare the intensity of FMR modes before and after OMCs attachment on the same MTJ. FMR studies were not planned around the time of magnetization study, and when all the experimental factors were in the optimized state. FMR data for the bare Ta/CoNiFe/AlOx/NiFe/Ta reported here was recorded from the different sample. In summary, OMCs produced single prominent (very weak) FMR mode after interacting with the bare Ta/CoNiFe/AlOx/NiFe/Ta (Ta/CoNiFe/AlOx/NiFe) MTJ with ferromagnetic (antiferromagnetic) inter-electrode coupling. The contrasting magnetization (Fig. 2 versus Fig. 5) and FMR (Fig. 12) studies suggested that OMCs enhanced the pre-existing coupling between two FM electrodes of a MTJ. Simultaneous evaluation of these results also supports that the only OMCs, not any artifact, is the reason behind the dramatic changes on MTJs.

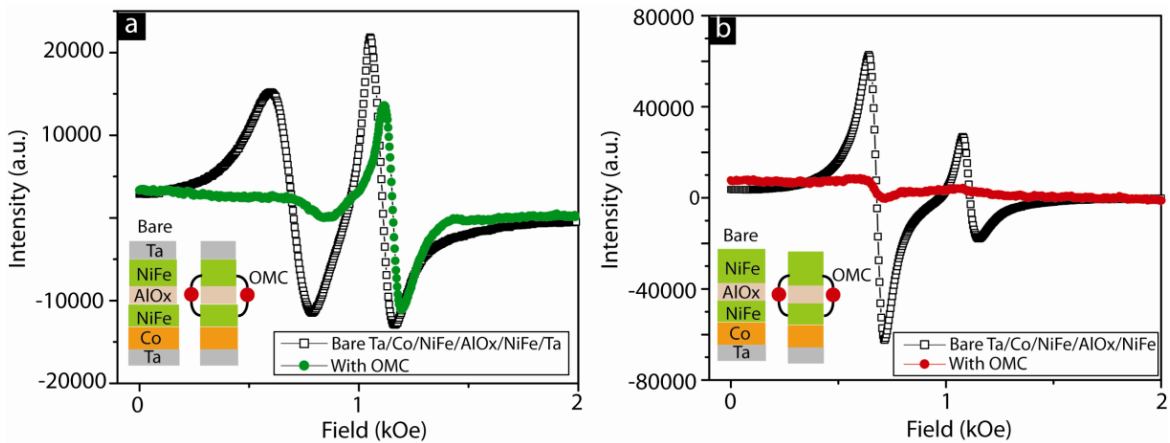

Fig. 12: Effect of OMCs on FMR modes of (a) Ta/Co/NiFe/AlOx/NiFe/Ta MTJ with weak ferromagnetic coupling, and (b) Ta/Co/NiFe/AlOx/NiFe MTJ with weak antiferromagnetic coupling.

Reversing the effect of molecules can produce an unequivocal support to the observation that only OMCs produced linear magnetization (Fig. 2a) and FMR mode suppression (Fig. 6). OMCs effect can be reversed by the two methods: (a) destroy the OMCs to recover the response of a bare MTJ or transform a MEMSD back to MTJ and (b) damage the tunnel barrier so that inter-electrode coupling via defects within the barrier supersede the OMCs effect, however in this case FMR response from damaged MTJ will be much different than that of a stable MTJ. We attempted both type of reverse experiments.

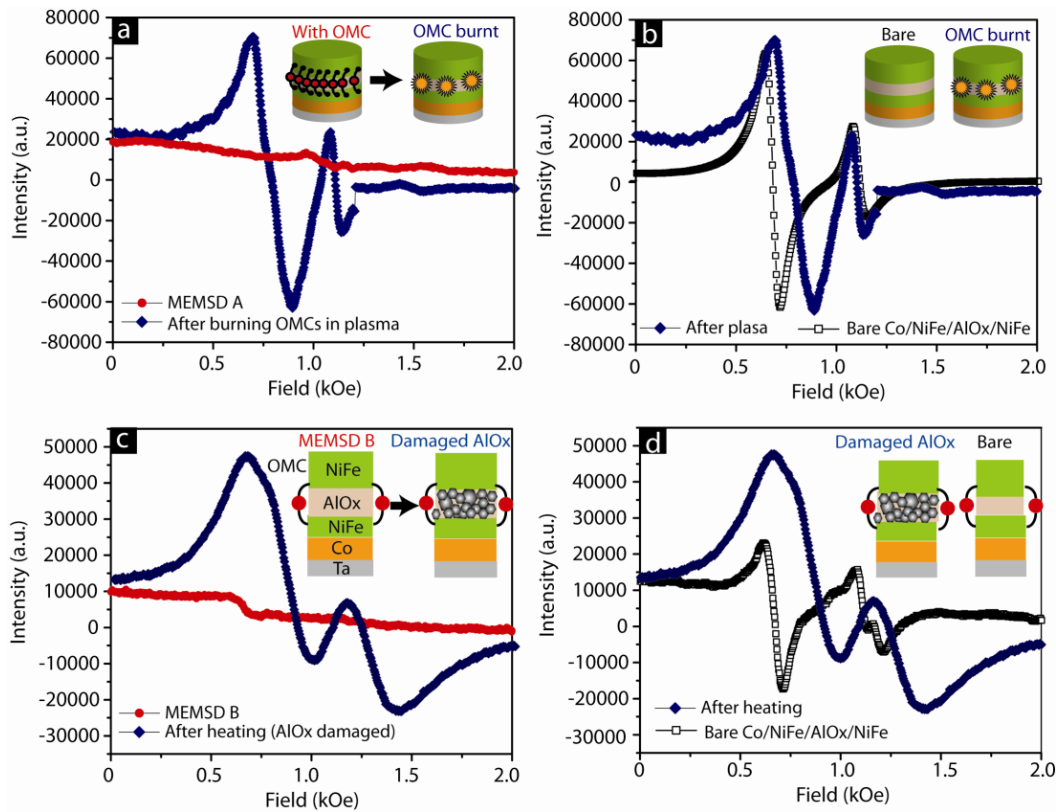

Fig. 13: Reversing the OMCs effect: (a) Burning the OMCs on MEMSDS-A in oxygen plasma produced two FMR modes (b) Two FMR modes appearing after plasma burning of OMCs are consistent with the typical two FMR modes corresponding to a bare MTJ. (c) Damaging AlOx barrier between, without affecting OMCs, produced diffused FMR modes, (d) Diffused FMR modes did not match with the modes of bare and stable MTJ.

In the first approach, OMC bridges were burnt in the oxygen plasma (inset of Fig. 13a-b). Oxygen plasma was generated by the 50 W RF bias at 60 mTorr pressure of 1:1 Ar:$O_2$ mixture in the sputtering machine chamber. For this study the MEMSD-A, on which magnetization and FMR study exhibited dramatic changes due to OMCs, was utilized. Prior to the burning of OMCs in oxygen plasma, MEMSD-A showed rather featureless FMR spectra (Fig. 13a). FMR study after burning the OMCs showed the emergence of two unambiguous FMR modes (Fig. 13a). The position and shape of these two modes are in excellent agreement with the acoustic and optical modes recorded on a separately grown bare Ta/Co/NiFe/AlOx/NiFe MTJ (Fig. 13b). It is noteworthy that we did not perform the FMR study on the MTJ before treating it with OMCs to produce MEMSD-A. Because MEMSD-A was fabricated for the magnetization study and at that time we did not plan for the FMR studies. The FMR spectra of plasma treated MEMSD-A and a bare MTJ are in excellent agreement and confirm following: (a) OMCs only transformed the FMR modes, (b) MTJ used for fabricating MEMSD-A possessed the same magnetic characteristics as observed with the typical bare MTJ.

In the second control experiment, we attempted to make OMCs induced coupling ineffective, but without damaging the OMC channels. Simple strategy we employed was based on creating stronger inter-electrode coupling via a planar ~ 2 nm AlOx barrier (inset of Fig. 13c-d). To create stronger coupling via defects, AlOx, tunnel barrier was damaged by thermal stresses [31]. MEMSD-B was heated at 390 K for 30 min in the flowing helium ambience, to avoid unintentional oxidation. Subsequently, this sample was cooled to room temperature and was subjected to FMR study. FMR spectra of this sample showed a broad FMR mode within 0-2 kOe range (Fig. 13c). We attributed the broadened resonance mode to the emergence of a wide range of exchange coupling strengths between the two FM electrodes via defective AlOx insulator [30]. It is noteworthy that the broad FMR mode and two distinct FMR modes of the bare MTJ utilized by MEMSD-B were in 0-2 kOe range (Fig. 13d). In essence, we were able to recover magnetic signal from the MEMSD, which suggested that dramatic reduction in FMR mode intensity was not due to the permanent structural damage of any kind.

MFM characterization was utilized as an approach to investigate the effect of OMCs on the individual MTJ. The role of MFM is crucial to support the results of magnetization and FMR studies which were unable to distinguish the response from the individual MEMSD. The magnetization and FMR studies exhibited the average response from several thousand MTJs and hence could not show the effect of OMCs from a single MTJ cylindrical dot. MFM images were produced by plotting the phase data over the scanned area. Bare MTJs (Ta/Co/NiFe/AlOx/NiFe) produced clear phase contrast and could be unambiguously recognized against nonmagnetic background like typical magnetic materials [23]. To understand the OMCs effect through MFM study, we mainly focused on MEMSD-B (Fig. 6), which showed OMCs induced modes suppression during FMR studies. Several MFM scans from regions of MEMSD-B showed negligible magnetic contrast at the sites of the MEMSDs. However, according to topographical scans the negligible MFM contrast was not due to the loss of MEMSDs at those sites (Fig. 14a). Topographical scans at the locations of MEMSDs produced unequivocal and uniform height data of ~34 nm, closely matching with the targeted physical thickness of the MEMSD cylindrical dot.

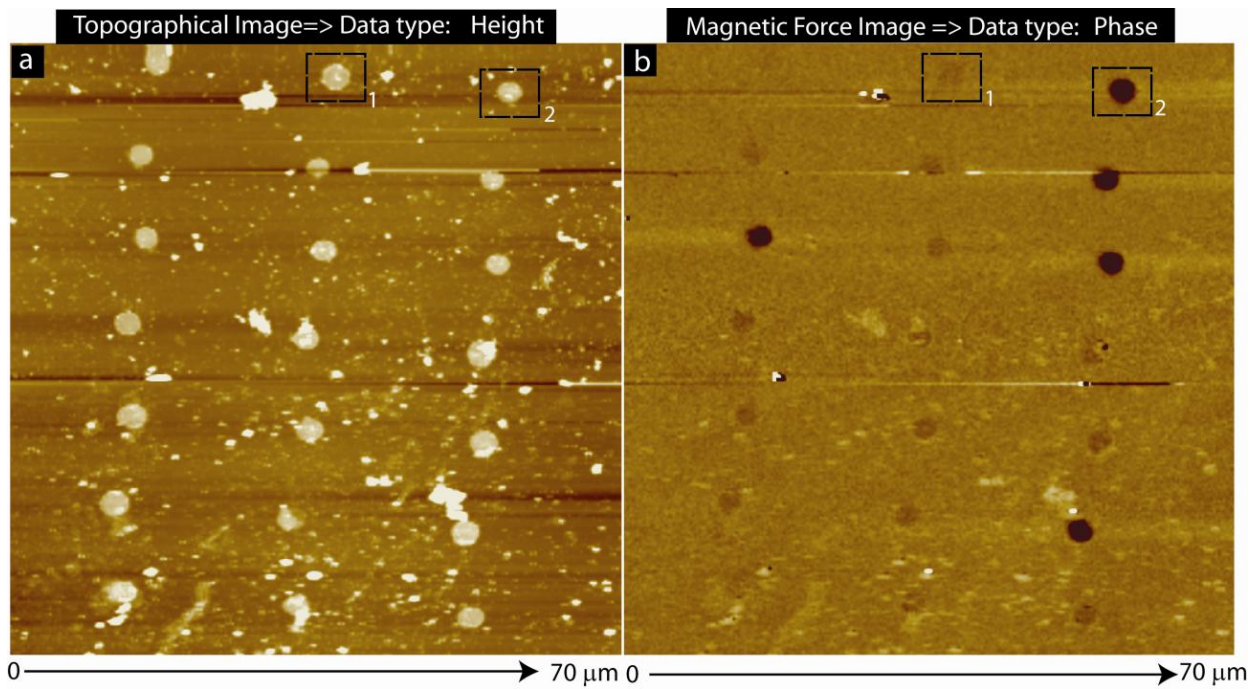

Fig. 14: MFM study of MEMSD-B: (a) Topographical image showing height data of MEMSDs and (b) corresponding magnetic force images were produced by using phase data.

Several self-referenced scans evidenced the accuracy of our MFM experimental procedure investigating the effect of OMCs. One can argue that source of low MFM contrast may be due to poorly optimized experimental factors. However, our MFM experimental data assert the accuracy of MFM procedure. Fortunately, in some MFM scans two types of magnetic force images were recorded (Fig. 14b). Majority of the MEMSDs cylinders showed highly dim magnetic contrast while some of the MEMSDs showed a clear magnetic contrast. High contrast MFM images resembled with that of the typical bare MTJs, and are believed to arise from those MTJs which were not affected by OMCs. We utilized high contrast phase images as the reference to discuss the majority of low contrast data (Fig.14b). The topographical image for the MEMSDs in box 1 and 2 are identical (Fig. 14a); the corresponding MFM images in box-1 and box-2 showed stark difference (Fig.14b). Such a dramatic change in two magnetic images was not possible if dim magnetic contrast was due to improper experimental procedure or any artifact. Furthermore, MFM images of three dots in the third row showed that central MEMSD dot is of the dim color contrast (Fig. 14a), while the two immediate neighbors showed high magnetic contrast (Fig. 14b). Again, such a dramatic change in magnetic images of the neighboring cylindrical dots is not possible if faint magnetic contrast was due to the improper experimental conditions. The simultaneous observation of dim and strong MFM contrast suggested that all the MTJs were not affected by the OMCs (Fig. 14b). This MFM scan showed that some MTJs remained unaffected by the OMCs and rationalize the FMR studies where OMCs showed partial effect (Fig.7c-f). Reduction in FMR signal intensity corresponded to the population of MTJ affected by OMCs (Fig. 7c-f).

Our magnetization study (Fig. 2a) rationalized the observed dim MFM contrast (Fig.14b) on MEMSDs. Magnetization study showed that Co/NiFe/AlOx/NiFe MTJs lost ~ 90% magnetic moments when treated with OMCs for producing MEMSDs, specifically MEMSD-A (Fig. 2a). In the simple term, the MFM contrast ($MFM_{contrast}$) of two antiferromagnetically coupled FM electrodes is presumably due to the difference in magnetic moments of the top ($M_{Top}$) and the bottom ($M_{Bot}$) FM electrodes (equation 3).

$$MFM_{Contrast} \propto |M_{Top} - M_{Bot}| \qquad (3)$$

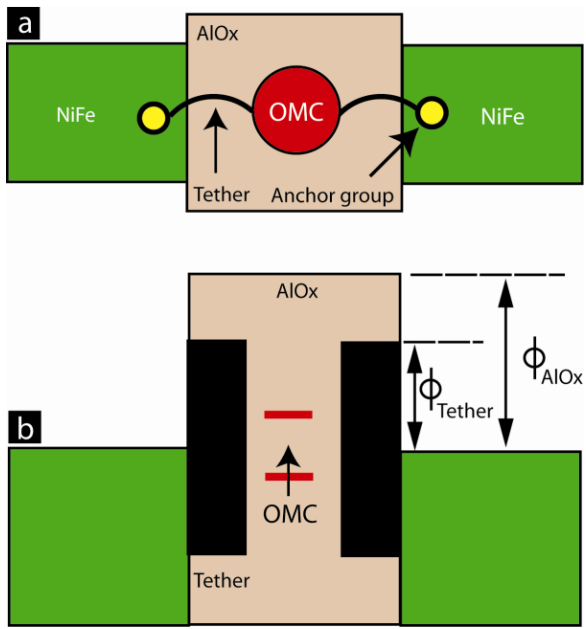

Fig. 15: (a) 2D image showing OMC channel bridging the AlOx insulator between two FM electrodes. (b) OMC modified energy band diagram showing additional molecular energy levels in between two FM electrodes and the modified barrier height.

Dramatic change in MFM contrast (Fig. 14) is in agreement with the equally dramatic changes in the FMR and magnetization studies. FMR and magnetization studies suggested that OMCs produced unprecedented strong AFC between the two FM electrodes of Co/NiFe/AlOx/NiFe MTJ. We hypothesized that OMCs not only produced a strong AFC but also served as a highly effective spin filter. Potential role of OMC as a spin filter can be justified by its chemical structure (Fig.1c). The center of an OMC is a cluster with a net spin, which is directly connected to FM electrodes via alkane spin channels. Alkane spin channels can preserve spin coherence for longer distance and time as compared to AlOx or inorganic mediums [14]. When an OMC become the dominant spin channel then it's core expectedly start working as a valve.

OMC's core will have the highest occupied molecular orbitals (HOMO) and the lowest unoccupied molecular orbitals (LUMO). It is surmised that HOMO and LUMO will allow the transport of only one type of spin via them. It means, HOMO and LUMO of OMC core will work as the two spin valves. This selective passage of one spin type through HOMO or LUMO siphon out one type of spin, e.g. spin up, and pump in opposite spins, e.g. spin down, at one of the two MEMSD's FM electrode. This spin filtering will produce two ~100% spin polarized FM electrodes. As a result in equilibrium state a MEMSD had two magnetic electrodes with equal density of states of the opposite spins, located at ~2 nm from each other. In MFM imaging magnetic probe was ~ 100 nm above the MEMSD. MFM probe interacted with the two electrodes of a MEMSD through the long range dipolar interaction. For this distant MFM probe net magnetic moment will be the sum of magnetic moment of the two OMCs modified FM electrodes. When two electrodes are antiparallel to each other, the magnetic moments of two electrodes cancel each other to give very faint MFM contrast (Fig. 14). This hypothesis is strongly supported by our magnetization studies, which established that OMCs reduced the effective magnetization of MTJ by ~90%. The magnetic moment reduction in our magnetization study (Fig. 2a) supports the reduction of the magnetic contrast in MFM study (Fig. 15).

Study of the mechanism behind an OMC's effect is crucial for the advancement of MEMSD. Following sections focuses on several perspectives useful in explaining the OMCs induced dramatic changes in the inter-electrode coupling. According to our magnetization study (Fig. 2a) and conventional definition [22] the lower bound of OMCs induced AFC strength was ~22 erg/cm$^2$ [Table-1]. This estimate showed that OMCs increased the coupling between two FM electrodes of a MTJ by ~5 orders. The typical inter-electrode coupling strength on a MTJ was < 1x10$^{-3}$ erg/cm$^2$ [9]. However, our calculation of J did not focus on several potentially important aspects like: (a) nature of interaction between a OMC and two individual FM electrodes, (b) role of biquadratic coupling, which is expected to emerge when two FM electrodes are coupled by nanoscale objects [28], (c) the role of spin fluctuations in enhancing overall exchange coupling, (d) the critical number of OMCs required to transform a ~10 μm$^2$ area MTJ with ~20 nm thick FM metals into a MEMSD.

Quantum mechanical calculations on such MEMSD configuration are extremely challenging. Recently, a few theoretical and experimental studies have been conducted to understand the magnetic interaction between magnetic molecules and a FM electrode [32]. The density functional theory (DFT) based studies were performed on a very small system involving single molecule and one FM electrode. Such calculations on a system of single molecule and few hundred FM atoms necessitated a large number of approximations [32]. Our MEMSDs involved ~10,000 OMCs simultaneously coupled with the two microscopic FM electrodes, and each FM electrode involved ~$10^{12}$ atoms. Though quantum mechanical calculations are critical to reveal the fundamental level understanding of MEMSD, but due to complexities and largeness of MEMSD such calculations are extremely challenging and beyond the scope of present work. For MEMSD it is more practical to utilize existing models and experimental studies to discuss the unprecedented coupling observed on MEMSDs.

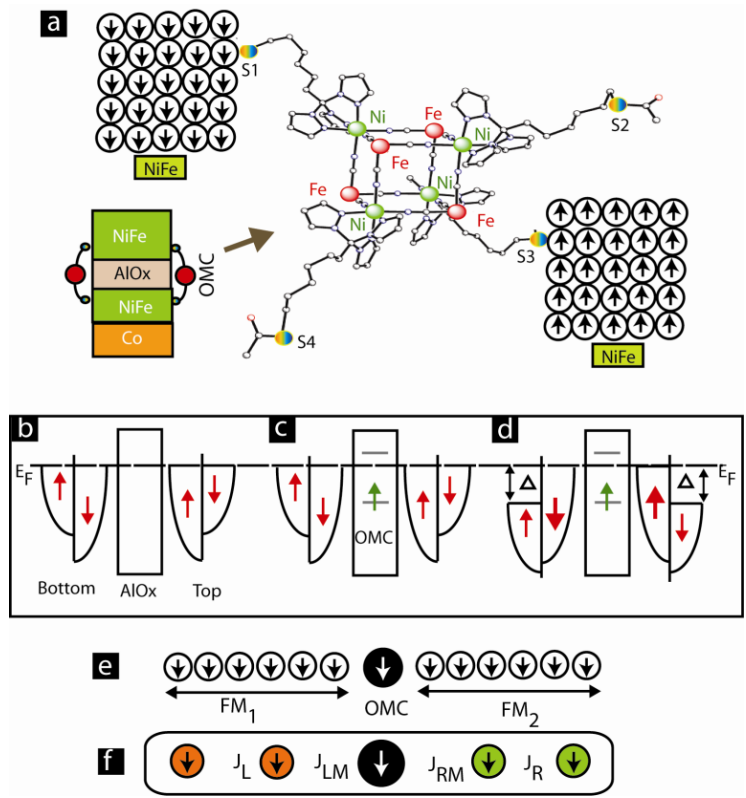

Fig. 16: MEMSD introduction: (a) Schematic showing bridging of single OMC between two NiFe electrodes. Band diagram of a MTJ (b) before and (c) after bridging OMC channels across insulator, as shown in the inset of (a). (d) New band diagram resulting from the OMC induced strong exchange coupling. (e) 1D analog of MEMSD and (f) display of exchange coupling strength between nearest neighbors of a OMC.

To understand the OMCs effect on MEMSD we focused on two aspects: (a) quantifying the magnitude of OMCs induced inter-electrode exchange coupling and (b) estimating the magnetic moment of OMC affected MTJ. For the quantification of inter- electrode exchange coupling OMCs can be viewed in two ways. In the first role, OMCs can be viewed as the agents that produce intermediate energy levels and modify tunnel barrier height and thickness (Fig. 15) [8]. OMCs are expected to produce an effective energy band diagram of the double barrier tunnel junction akin to quantum wells [33]. After the establishment of OMC channels (Fig. 15a) effective barrier height changed from $\Phi_{AlOx}$ to $\Phi_{OMC}$ and effective barrier thickness reduced from physical thickness of AlOx to the physical length of tether attached to OMCs. To quantify OMCs induced coupling we have utilized Slonczewski's model derived for MTJs [34]; we rationalize the selection of this model (equation 4) because our MEMSD are the modified MTJs or customized MTJs.

$$J = \frac{\phi_B k^3 (k^2 - k_\uparrow k_\downarrow)(k_\downarrow - k_\downarrow)^2 (k_\downarrow + k_\uparrow)}{\pi^2 d^2 (k^2 + k_\uparrow^2)^2 (k^2 + k_\downarrow^2)^2} e^{-2k.d} \qquad (4)$$

Here, d and $\Phi_B$ are the barrier thickness and barrier height, respectively, of the insulating medium between two spin-polarized electrodes. Whereas, $k_\downarrow$ and $k_\uparrow$, are the wave vectors of the spin-down and spin-up electrons of the FM electrodes. *Equation 4 couple the electrons' wave vectors in the FM electrode with the electrons' wave vector in the region between FM electrodes.* The wave vector of an electron in the insulating regime is k and is given by k=√(2.$\phi_B$.m.$\hbar^{-2}$). Here, m is the effective mass of the electron and $\hbar$ is Planck's constant. We assume that a OMC, which has net spin state, behaves like a spin-polarized electrode and hence reduces the effective barrier thickness from 2 nm (AlOx thickness) to 1 nm (distance between center of the OMC core to the FM electrode edge). However, we are unable to determine $k_\downarrow$ and $k_\uparrow$ for the FM electrodes. $k_\downarrow$ and $k_\uparrow$ strongly depends on the interfacial properties and the spacer(s) between the two FM electrodes. According to our magnetization, FMR (Fig. 6), and MFM studies, OMCs produced dramatic changes in the magnetic properties of FM electrodes. The calculation of $k_\downarrow$ and $k_\uparrow$ for OMCs affected FM electrodes is challenging and beyond the scope of present work. We rather choose to work with the following simplified form of equation 4 [34] given in equation 5.

$$J \propto \frac{\exp(-2k.d)}{d^2} \qquad (5)$$

According to the analysis of charge transport data, barrier thickness (d) changed from ~2 nm to ~1 nm and barrier height changed from ~1 eV to 0.6 eV [8] after OMCs attachment. These OMCs induced modifications in barrier height and barrier thickness increased the magnitude of J by ~5 orders.

This calculation does not explain utility of OMCs; it is arguable that such change in barrier height and thickness can also be produced by simply reducing the barrier thickness using a thinner insulator or reducing the barrier height by using any appropriate insulating material. However, to date manipulation of the insulating barrier between FM electrodes have not yielded dramatic changes according to the one reported here. However, modifying barrier properties in a trilayer thin film configuration with film thickness in few monolayer ranges exhibited intriguing effect of inter-electrode exchange coupling. In a related study, reducing the thickness of *non-magnetic metallic spacer* between the two FM electrodes to few monolayers produced dramatic changes in the Curie temperature of the constituents FM electrodes [18, 35]. Importantly, this phenomenon could not be explained in the conventional theoretical frame work as utilized for the development of Slonczewski's model derived for MTJs [34] (equation 4); additional consideration of role of spin fluctuations satisfactorily explained the dramatic change in the basic magnetic properties like Curie temperature of the FM electrodes [18]. Spin fluctuations are also expected to be crucial for defining the anomalous magnetic response of MEMSD type molecular spin devise as well. In fact theoretical and experimental studies on devices akin to MEMSD have established the utility of spin fluctuations. In the theoretical study by Martinek et al. [36], spin fluctuations assisted coupling produced anomalous Kondo level splitting phenomenon when a magnetic quantum dot or nanoscale object was inserted between the two FM electrodes. In an experimental study, Pasupathy et al. [5] realized the spin fluctuation assisted anomalous Kondo splitting on a molecular spin device configuration. This molecular spin device was produced by placing a $C_{60}$ molecule with net spin state in the nanogap of nickel breakjunction. However, Pasupathy et al. [5] did not report any exclusive magnetic characterization like magnetization, FMR, and MFM before and after inserting $C_{60}$ molecules in the Ni breakjunction. Because of the lack of magnetic studies, it is uncertain if $C_{60}$ induced strong inter-electrode coupling also produced dramatic changes in the magnetic properties of Ni FM electrodes, as observed on our MEMSDs when OMCs were bridged between the two FM electrodes (Fig. 1).

To quantify the magnitude of OMCs induced exchange coupling, we investigated other analogous systems. In the first comparison, we considered OMCs to be analogous to atomic impurities. Zhuralev et al. [9] have studied the nature and strength of impurities enhanced exchange coupling between the two FM electrodes of a MTJ. This theoretical study provide satisfactory explanation to the experimental observation of impurities enhanced exchange coupling and the magnitude of inter-electrode coupling [9, 21]. However, the magnitude of theoretically calculated inter-electrode exchange coupling was ~ 2 orders smaller than he magnitude of lower bound of OMCs induced inter-electrode exchange coupling on our MEMSDs (Table-1).

In the second comparison, we considered OMCs to be analogous to magnetic nanoclusters. Wong et al. [10] studied the effect of iron (Fe) nanoclusters on the inter electrodes exchange coupling of a MTJ. In this study, Fe nanoclusters were embedded within the magnesium oxide (MgO) tunnel barrier, throughout the planar area of a MTJ. The MTJ in this study included one hard and one soft FM electrodes, similar to our MEMSD configuration. Depending on the Fe nanoclusters position in the MgO tunnel barrier, the inter-electrode exchange coupling not only exhibited a significant variations in magnitude but also switched nature of coupling between antiferromagnetic and ferromagnetic types [10]. Nanoclusters, however, only enhanced the inter-electrode coupling by a factor of < 2. One of the most important finding of this study is the unambiguous demonstration of the effect of using hard and soft FM electrodes for observing the strongest effect of nanoclusters. Nanoclusters induced inter-electrode exchange coupling was 160% higher for Fe/MgO/Co (MTJ with hard and soft FM electrodes) than that of Fe/MgO/Fe (MTJ with soft FM electrodes). *This study [10] complement our observation of the strongest OMC effect when MEMSD employed the MTJs with hard and soft FM electrodes only.* The observation of current suppression phenomenon discussed elsewhere, and the observation of change in magnetic properties discussed in the present paper required Co/NiFe/AlOx/NiFe MTJ. In a MEMSD, the Co/NiFe electrode was ~4 times magnetically harder than the NiFe electrode. In addition to our MEMSD, Pasupathy et al.'s [5] experimental observation of anomalous Kondo splitting also required a molecular spin device where two FM electrodes possessed significantly different magnetic hardness.

The interaction of nanoscale objects with the FM electrodes of different magnetic hardness governs the strength of inter-electrode exchange coupling (J). Wong et al. [10] satisfactorily modeled the effect of nanoclusters by focusing on its interaction with the individual FM electrodes. We customized the Wong et al. [10] model to express the OMC induced cumulative J in a MEMSD (equation 6):

$$J = J_{AlOx} + \frac{\left|J_{Soft-OMC}+J_{Hard-OMC}\right|-\left|J_{Soft-OMC}-J_{Hard-OMC}\right|}{2} \tag{6}$$

Here, $J_{Soft-OMC}$ ($J_{Hard-OMC}$) is the exchange coupling between OMC and soft (hard) FM electrode, and $J_{AlOx}$ is the inter-electrode exchange coupling via ~ 2 nm AlOx insulator. In the present case $J_{AlOx} << J_{Soft-OMC}$ or $J_{Hard-OMC}$. Assuming, $J_{Soft-OMC} \approx J_{Hard-OMC}$ and using the estimate of OMCs induced J from table 1 we found that $J_{Soft-OMC}$ or $J_{Hard-OMC}$ is of the order of ~22 erg/cm$^2$. The OMCs induced exchange coupling is 3-4 orders higher than the exchange coupling produced by the nanoclusters.

In essence, OMC produced much stronger effect as compared to impurities and nanoclusters. OMCs' strong effect is mainly due to its uniform shape and size that is not possible with nanoclusters and impurities. Moreover, nanoclusters are separated from the FM electrodes by MgO and represent the case of physiosorption interaction; whereas OMCs are covalently attached to FM electrodes via covalent bonding through thio functional group (Fig. 16a). The nature of molecule- metal bonding strongly affect the molecular transport and inter-electrode coupling [37]. For instance, the cobalt phthalocyanine (CoPc) molecule exhibited ~ 200 K Kondo temperatures when the nature of molecule-metal interaction changed from physiosorption to covalent bonding [37]. The nature of OMC interaction with FM electrodes of MEMSD will critically affect the effective exchange coupling [32]. Accurate estimation of J of MEMSD will also necessitate the experimental studies focusing on the determination of OMCs interaction with individual FM electrodes.

After the discussion on the magnitude of exchange coupling it is imperative to focus on the mechanism by which a strong exchange coupling can affect the magnetic properties of MEMSD's FM electrodes. Our magnetization, FMR and MFM studies showed dramatic changes in the inter-electrode coupling and the magnetic moment of the FM electrodes when a MTJ was treated with OMCs. To elucidate the effect of OMC bridges we compared our MEMSD with Ni brekjunction based molecular spin device [5]. In the Ni break junction, one or few $C_{60}$ molecules connected the two Ni FM electrodes similar to the case of MEMSD where ~10,000 OMCs connected the Co/NiFe and NiFe FM electrodes. Pashupaty et al. [5] reported that $C_{60}$ molecule(s) invoked spin fluctuation assisted strong exchange coupling. This strong exchange coupling produced a local magnetic field, which influenced the correlated Kondo energy level arising due to the strong overlap between the molecule and metallic energy levels. *This strong coupling induced local magnetic field is equivalent to the external magnetic field.* We expect that in our case an OMC produced similar magnetic coupling induced local magnetic field as produced by a $C_{60}$ molecule [5]. Pasupathy et al. [5] estimated that only one or few $C_{60}$ molecules produced > 50 T local magnetic field. In our case, a much stronger local magnetic field is expected since ~10,000 OMCs are coupling the FM electrodes of a MEMSD. We hypothesized that OMC induced local magnetic field will influence the stoner splitting of FM electrodes. *The Stoner splitting determines the energy gap between DOS of spin up and spin down electrons.* The local magnetic field will influence the stoner splitting in the same manner as an external magnetic field will influence the Zeeman splitting. Molecule induced local magnetic field in turn can produce a difference in chemical potential under the effect of magnetic field to yield much different spin DOS and different spin polarization (P). In addition, OMC's HOMO-LUMO is expected to serve as a spin filter to amass only one type of spin on each of the MEMSD's FM electrode. A plausible mechanism about the OMC dependent spin filtering has been discussed in the context of MFM studies of MEMSD (Fig. 14) In order to determine the magnitude of OMCs induced local magnetic field we have customized the model used by Pasupathy et al. [5]. In the context of MEMSD with two anti-parallel FM electrodes the coupling induced magnetic field (B) can be given by the following model (equation 7):

$$B \approx \frac{a \sum_{n=1}^{N}(P_T \Gamma_{nT} - P_B \Gamma_{nB})}{g \cdot \mu_B} \quad (7)$$

Here, N is the number of OMCs bridging the gap between FM electrodes, $P_T$ ($P_B$) is the degree of spin polarization of the top (bottom) FM electrode. $\Gamma_{nT}$ ($\Gamma_{nB}$) is the coupling energy of the $n^{th}$ molecule with the top (bottom) layer. g and $\mu_B$ denoted the gyromagnetic ratio for the molecule and Bohr magnetron, respectively.

In order to estimate the magnitude of B, we can assume N=10,000, aP is 0.15 [26] and $g\mu_B$ ~115 µV/T [38]. Generally molecular coupling energy is ~10 meV [5]. Without delving into coupling strength of an OMCs with the top and bottom FM electrode, we focus on the difference in coupling energy. Assuming the difference in coupling energy is 0.001-0.01 mV, we obtained B in the range of ~100-1000 T. We assumed, this OMC induced local magnetic field is applied on the FM electrodes to produce Stoner splitting and can be quantified by the following expression [39] (equation 8).

$\Delta = \pm 0.5(Pg\mu_0\mu_B H)$                                 (8)

Assuming P~0.5 for NiFe [1], the magnitude of OMC induced additional Stoner splitting was found to be in the 30-300 mV range. This analysis shows that MEMSD's FM electrodes can indeed undergo reshuffling of DOS (Fig. 16b-d), which in turn will affect their degree of spin polarization. However, this analysis by no mean is the best one, and would need significant adjustments to produce a more realistic estimate. The FM atoms, which are in direct contact of an OMC, are likely to experience the strongest effect. However, our magnetic studies have shown that OMCs induced exchange coupling influenced the microscopic junction area, not just few atoms in the vicinity of molecular junctions. It means there must be some mechanism in place to transport the OMC coupling effect from its nearest neighboring FM atoms. We surmise that Heisenberg exchange coupling and dipolar long range coupling among the atoms of a FM electrode transpire the OMC's coupling effect into the regions, which are even several μm off from the molecular junction. In order to validate our hypothesis we set up 1D Monte Carlo simulations on the system representing 1D form of a MEMSD (Fig. 16e). The magnetic ordering of FM electrodes (FM1 and FM2) was studied with the variation of exchange coupling strength of OMC with the FM electrodes. During Monte Carlo simulation, we minimized the system energy as given by the following expression (equation 9):

$$E = -\sum_{i=1}^{N} J_L S_i S_{i-1} - \sum_{i=1}^{N} J_R S_i S_{i-1} - J_{LM} S_L S_M - J_{RM} S_R S_M$$

In the above equation: $J_L$, $J_R$, $J_{LM}$, and $J_{RM}$ are the Heisenberg exchange coupling strengths for the FM electrode on the right, FM electrode on the left, between the left metal and the molecule, between the right metal and the molecule, respectively (Fig. 16f). $S_i$ and $S_{i-1}$ are the spins of the nearest neighbors. $S_L$ and $S_R$ are the spins of left and right FM electrode's atoms, which are the nearest neighbor to the molecule. Initially, we considered only the Heisenberg interaction, and employed a periodic boundary condition [40]. Usage of the periodic boundary condition ensured that the spins on the one edge of the Ising lattice are the nearest neighbor to the corresponding spin on the opposite edge [40]. After choosing appropriate values for the Heisenberg exchange coupling coefficient, temperature and random spin states, we set up a Markov process to generate new states. Under the Metropolis algorithm, the spin of a randomly selected site was flipped to produce a new state. New state was only rejected if difference between the final and new energy (ΔE) satisfied the following condition (equation 10):

$$\{\Delta E > 0, \ \exp(-\Delta E / kT) < rand\# \tag{10}$$

Following these strategies, we generated new configurations of 1D MEMSD, and calculated the magnitude of observables. First, we studied the magnitude and sign of OMC-FM electrode exchange coupling required to observe the long range OMC's effect. According to Monte Carlo simulations, the following two conditions about the sign and magnitude of OMC-FM electrode exchange coupling were critical to observe the experimentally observed OMC effect. According to the first condition OMC's exchange coupling with the two FM electrodes must be of the opposite sign. According to the second condition OMC exchange coupling with at least one FM electrode was so that the $J_{LM}/J_L$ (or $J_{RM}/J_R$) ratio was in the range of 0.6 to 3.5. If the $J_{LM}/J_L$ (or $J_{RM}/J_R$) ratio was <0.6 or >3.5 then only unstable ordering occurred on 1D MEMSD. Here, the requirement of $J_{LM}/J_L$ (or $J_{RM}/J_R$) ratio to be >0.6 suggested that for long range ordering, the magnitude of OMC exchange coupling strength should be at least 60% of the magnitude of exchange coupling strength between the two atoms of the left (or right) FM electrode. We studied 1D MEMSD system up to 1000 FM atoms, each FM electrodes having 500 atoms (Fig. 16e). In conclusion, magnetic coupling with one OMC remained effective in maintaining antiparallel arrangement between the two FM electrodes.

Our Monte Carlo studies provided deeper insight about the exchange coupling of an OMC with the two FM electrodes. Interestingly, the magnitude of OMC's exchange coupling with the two FM electrodes can be very different. For instance if $J_{RM}/J_R$ was in 0.6 to 3.5 range, then $J_{LM}/J_L$ could be as small as -0.01 to produce a long range antiferromagnetic ordering on to the two FM electrodes. This simulation result is consistent with the experimental observation of significant change in magnetic moment of MTJ with one FM electrode. These MTJs possessed Ta/Co/NiFe/AlOx/Pd and NiFe/AlOx/Pd configurations. Our magnetization studies with these MTJs represented the case when OMC exhibited a strong magnetic coupling with one electrode (FM electrode) and weakly connected to another electrode (nonmagnetic or may be weakly magnetic Pd electrode [41]. Pd has exhibited ferromagnetism when treated with thiol ended molecules like the one used in the present study [8, 19]; hence Pd may be considered as weak ferromagnet whose couling strength with OMC is expectedly weak. OMCs increased (decreased) the magnetic moment of the MTJ with Ta/Co/NiFe (NiFe) magnetic electrode (Fig. 4). In summary, Monte Carlo simulation and magnetization studies suggested that a OMC established ferromagnetic and antiferromagnetic exchange coupling with the Ta/Co/NiFe and NiFe FM electrodes, respectively.

1D MEMSD simulations were also used for estimating the magnitude of OMC induced inter-electrode coupling strength. According to the model developed by Wong et al. [10] the OMC induced exchange coupling depended on the magnitude of an OMC exchange coupling with the two FM electrodes (equation 6). Assuming, $J_{LM}/J_L = J_{RM}/J_R = 0.6$, $J_R = J_L = J_{NiFe}$ and $J_{LM}$ and $J_{RM} \gg J_{AlOx}$, the strength of molecule induced exchange coupling (J) is = 0.6 $J_{NiFe}$. Here, $J_{NiFe}$ is the interatomic exchange coupling strength in NiFe FM electrode. Interatomic exchange coupling strength of a ferromagnetic material is equivalent to its Curie temperature; Curie temperature governs the amount of thermal energy required to disrupt the ferromagnetic ordering due to inter-atomic ferromagnetic exchange coupling. Hence, $J_{NiFe}$ can be directly estimated from the Curie temperature of NiFe metal. It is noteworthy that an OMC is only connected to NiFe electrode; hence utilization of Curie temperature for the NiFe will provide the order of its exchange coupling ($J_{NiFe}$). Utilizing NiFe's Curie temperature to be ~900 K [23, 28] the OMC induced exchange coupling turned out to be ~540 K.

The magnitude of OMC induced exchange coupling from Monte Carlo simulation is in agreement with the experimentally observed exchange coupling in our magnetization study. Magnetic moment versus temperature study of MEMSD-A exhibited a sharp transition corresponding to ~330 K Neel temperature (Fig. 3). Neel temperature of MEMSD represented the magnitude of OMCs induced AFC between the two FM electrodes. Antiferromagnetic surface oxides of NiFe can also exhibit similar transition; however we argue against this possibility because Neel temperature of the oxides of NiFe and Co are much higher than ~330 K, the temperature which is believed to represent OMC induced exchange coupling strength. Neel temperature of MEMSD is closer to the magnitude of theoretically calculated coupling strength from our 1D MEMSD Monte Carlo simulations. For more accurate analysis similar MC simulations with higher dimensionality and inclusion of the other potential factors like dipolar coupling, biquadratic coupling, and magnetic anisotropies of the individual FM electrodes will be required.

**Conclusion:**

Maneuvering inter-layer coupling has strong potential to produce novel magnetic materials out of existing magnetic materials. Here we discussed MEMSD system in which molecular coupler produced dramatic effect on the magnetic properties of the ferromagnetic electrodes. MEMSD paradigm offers enormous flexibility and opportunities to develop novel magnetic metamaterials. MEMSD can produce spin valves and molecular spin devices for the application in quantum computation. MEMSD is also suitable for fundamental level understanding because of the possibility of performing extensive transport and magnetic studies. Present work exhibited a strategy to control the inter-ferromagnetic electrode coupling by the molecular coupler. This work also encourage the exploration of similar observation with interesting combinations of magnetic molecules like single molecular magnets and the various configurations of magnetic tunnel junctions.

**Acknowledgments:**

PT thanks Prof. Bruce J. Hinds and the Department of Chemical and Materials Engineering, University of Kentucky to enable his PhD research work presented in this manuscript. He also thanks D.F Li and S. M. Holmes for providing molecules used in this work.